\documentstyle[A4,12pt]{article}

\begin{document}

\bibliographystyle{plain}
%%%%%%%%%%%%%%%%%%%%%%%%%%%%%%%%%%%%
%---------REDEFINITIONS-------------
\renewcommand{\theequation}{\arabic{section}.\arabic{equation}}
\renewcommand{\arraystretch}{1.2}
\newcommand{\beq}{\begin{equation}}
\newcommand{\eeq}{\end{equation}}

\newcommand{\bea}{\begin{eqnarray}}
\newcommand{\eea}{\end{eqnarray}}
\newcommand{\tr}{\mbox{\rm tr}}

\newcommand{\bean}{\begin{eqnarray*}}
\newcommand{\eean}{\end{eqnarray*}}

\newcommand{\bei}{\begin{itemize}}
\newcommand{\eei}{\end{itemize}}
%---------DEFINITIONS---------------
\def\esp{\mbox{\hspace{10pt}and\hspace{10pt}}}
\def\gm{g^{-1}}
\def\gl{g_{L}}
\def\jl{j_{L}}
\def\Jl{J_{L}}
\def\ml{\gamma_{L}}
\def\gr{g_{R}}
\def\jr{j_{R}}
\def\Jr{J_{R}}
\def\mr{\gamma_{R}}
\def\xp{x^{++}}
\def\xm{x^{--}}
\def\allie{{\cal G}}
\def\rc{{r_{c}}}
\def\rpm{r^{\pm}}
\def\rmt{{r_{m}}}
\def\rpmt{r^{\pm}_{m}}
\def\dd{{\cal D}}
\def\ddd{\bar{\cal D}}
\def\jb{\bar J}
\def\dexyz{\delta^{0}_{xy}}
\def\dexyu{\delta^{1}_{\mathbf{xy}}}
\def\dexyd{\delta^{2}_{XY}}
\def\bthpz{\Theta^{0}(x-y)}
\def\bthpu{\Theta^{1}(\mathbf{x-y})}
\def\bthpd{\Theta^{2}(X-Y)}
\def\bthnz{\Theta^{0}(y-x)}
\def\bthnu{\Theta^{1}(y-x)}
\def\bthnd{\Theta^{2}(Y-X)}
\def\thpz{\Theta^{0}_{xy}}
\def\thpu{\Theta^{1}_{\mathbf{xy}}}
\def\thpd{\Theta^{2}_{XY}}
\def\thnz{\Theta^{0}_{yx}}
\def\thnu{\Theta^{1}_{yx}}
\def\thnd{\Theta^{2}_{YX}}
\def\bepz{\varepsilon^{0}(x-y)}
\def\bepu{\varepsilon^{1}(\mathbf{x-y})}
\def\bepd{\varepsilon^{2}(X-Y)}
\def\epz{\varepsilon^{0}_{xy}}
\def\epu{\varepsilon^{1}_{\mathbf{xy}}}
\def\epd{\varepsilon^{2}_{XY}}
\def\benz{\varepsilon^{0}(y-x)}
\def\benu{\varepsilon^{1}(y-x)}
\def\bend{\varepsilon^{2}(Y-X)}
\def\enz{\varepsilon^{0}_{yx}}
\def\enu{\varepsilon^{1}_{\mathbf{yx}}}
\def\enpd{\varepsilon^{2}_{YX}}
\def\gg{g_{1}(x)g_{2}(y)}
\def\GG{G_{1}(X)G_{2}(Y)}
%--------THEOREM--------------
\newtheorem{prop}{Property}
\newtheorem{remark}{Remark}
\newtheorem{defi}{Definition}
\newtheorem{theo}{Theorem}
%---------COMMANDS------------------
\newcommand{\pcc}[2]
	{
%\noident
\{ {#1}_{1}(x), {#2}_{2}(y) \}
	}

\newcommand{\dh} [1]
	{
\partial^{{#1}} g
	}

\newcommand{\Pcc}[2]
	{
%\noident
\{ {#1}_{1}(X), {#2}_{2}(Y) \}
	}

\newcommand{\db} [1]
	{
\partial_{{#1}} g
	}

\newcommand{\reff} [1]
	{
(\ref{#1})
	}
%%%%%%%%%%%%%%%%%%%%%%%%%%%%%%%%%%%%

\renewcommand{\thefootnote}{\alph{footnote})}
\newcommand{\un}{\underline}
\oddsidemargin=0.25cm\evensidemargin=0.25cm
\thispagestyle{empty}
\begin{flushright}ENSLAPP-L-560\\
hep-th/9512221 \\
December 1995
\end{flushright}
\vskip 1.0truecm
\begin{center}{\bf\Large
$N=2$ chiral WZNW model in superspace}
\end{center}
 \vskip 1.0truecm
\centerline{\bf F. Delduc, M. Magro}
\vskip 1.0truecm
\centerline{ \it Lab. de Phys. Th\'eor. ENSLAPP
\footnote{URA 14-36 du CNRS, associ\'ee  \`a l'ENS de Lyon et au LAPP}
, ENS Lyon}
\centerline{\it 46 All\'ee d'Italie, 69364 Lyon, France}
\vskip 2.0truecm  \nopagebreak

\begin{abstract}
We study the  Poisson bracket algebra of the $N=2$ supersymmetric chiral
WZNW model in superspace. It involves two classical r-matrices, one of which
comes from the geometrical constraints
implied by $N=2$ supersymmetry. The phase space itself consists of superfields
satisfying constraints involving this r-matrix. An attempt is made to relax
these constraints. The symmetries of the model are investigated.
\end{abstract}
\setcounter{page}0
\newpage

\section*{Introduction}

It is well-known that the exchange algebra of the chiral WZNW model involves a
classical $r$-matrix \cite{blok89,fadd90,alek90}. This model possesses a
Poisson-Lie symmetry which is the classical counterpart of the quantum group
symmetries \cite{falc91,falc93,gawe91,alek92,chu91}. In this paper we shall
consider the $N=2$ supersymmetric extension of this model.

In contrast with the $N=1$ supersymmetric extension of a non-linear
$\sigma$-model, the $N=2$ extension requires some geometrical properties of the
target manifold. These conditions were found in \cite{alva81} for target spaces
without torsion, then in \cite{gates,howe,hull85,hull87} for target spaces with
torsion and specialized in \cite{sevr88-1,sevr88-2,sevr88-3} for the case of
parallelized group manifolds. In this last case, these conditions were recently
reformulated in terms of Manin triples \cite{park92,getz93-2,figu94}. In
particular the existence of a $N=2$ WZNW model on $G$ requires the existence of
two Manin triples induced by two $r$-matrices.

The geometrical properties of supersymmetric models are usually more
transparent when formulated in the appropriate superspace, where the invariance
under supersymmetry is explicit. Formulating the $N=2$ WZNW model in superspace
\cite{hull90-3} leads to constrain each supercurrent to belong to one of the
isotropic subalgebras present in the Manin triples. In this article, we focus
on the study in superspace of the chiral WZNW model. The phase space variables
are constrained $(2,0)$ superfields taking value in the group $G$, together
with the usual monodromy. The Poisson bracket involves the r-matrix of the
bosonic exchange algebra, together with the r-matrix defining the Manin triple.
Only this last r-matrix is present in the Poisson brackets of the currents,
which are
shown to coincide with the classical limit of \cite{hull90-2}.
Although this Poisson bracket is compatible with the constraints on the
N=2 superfields, the Jacobi identity  works without any use of these
constraints.
It is thus possible to consider an enlarged phase space made of general
(2,0) superfields. We construct an energy-momentum tensor which Poisson bracket
form, both in the constrained and non-constrained cases, a classical $N=2$
super-Virasoro algebra. However, only in the constrained case do the field
equations reduce to chirality conditions.

A consequence of the constraints is that the $N=2$ WZNW model is not invariant
under the whole of $N=2$ loop group transformations. We compute a convenient
form of the infinitesimal loop transformations consistent with the constraints.
The Grassman coordinates of the superspace appear explicitly in their
expression, and thus they are not superfields \cite{hull90-3}. The moment map
corresponding to these transformations is given.

In \cite{hull90-3,hull90} it was proposed to restore the commutation of these
transformations by adding a compensating transformation of the $r$-matrix which
appear in the Manin triple. This procedure is reminiscent of the action of a
Poisson-Lie group (such groups are closely related to $r$-matrices) on a
Poisson manifold. Such an action  is a symmetry only because the parameters of
the transformation carry their own Poisson bracket. This suggests that one
could perhaps consider all $N=2$ loop group transformations as symmetries if
they act as Poisson-Lie transformations. However, since these transformations
do not preserve the constraints, we have to consider the unconstrained phase
space. Indeed there is in this phase space a Poisson-Lie action of $N=2$
loop transformations. But since we do not have a clear understanding of the
field equations there, this program cannot be considered as completed.

This paper is organized as follows. In section \ref{eq99} and \ref{eq105} we
review the WZNW model as a non-linear $\sigma$-model, and its chiral
counterpart. In section \ref{eq100} we recall the properties of the $N=1$
extension of these models using $N=1$ superspace and superfields. In section
\ref{eq101} we recall how the existence of an $N=2$ extension is related to the
existence of a Manin triple. Then in section \ref{eq102} we formulate the
equations of the model in $N=2$ superspace and compute the exchange algebra of
the chiral $N=2$ WZNW model in section \ref{eq103}. The appendix contains the
definitions and some properties of the extended  supersymmetric Dirac and sign
distributions.
\section{The WZNW model}
\setcounter{equation}{0}
\label{eq99}
	\subsection{The WZNW model as a non-linear $\sigma$-model}
\label{eq106}
		The two-dimensional WZNW model \cite{WittWZW} is defined by the action
\beq
S[g] \equiv -(8\pi)^{-1}k \int_{M} \tr(\gm\db{\mu})(\gm\dh{\mu})dx^{0}dx^{1} +
(12\pi)^{-1}k \int d^{-1} \tr(\gm dg)^{3} \label{eq2}
\eeq
where $g$ is a map from the cylinder $M$ to a semi-simple Lie group manifold
$G$ with corresponding Lie algebra ${\cal G}$,
\beq
g(x^0,x^1+2\pi)=g(x^0,x^1).
\eeq
The light-cone coordinates of the world-sheet $M$ will be denoted by
$x^{\pm\pm} = x^{0} \pm x^{1}$ where the index refers to the Lorentz weight
carried by the coordinate. The corresponding derivatives read
$\partial_{\pm\pm} \equiv \frac{1}{2}(\partial_{0} \pm \partial_{1})$. When
written in terms of local coordinates $\varphi^{i}$ on $G$, the action
(\ref{eq2}) takes the usual non-linear $\sigma$-model form
\beq
S[\varphi] \equiv \int_{M} dx^{++}dx^{--}(g_{ij} + b_{ij})(\varphi)
\partial_{++}\varphi^{i}\partial_{--}\varphi^{j}, \label{eq71}
\eeq
where $g_{ij}$ and ${\cal T}_{ij}^{k} \equiv \frac{1}{2}
g^{kl}(b_{il,j}+b_{lj,i}+b_{ji,l})$ are respectively the metric and the torsion
on $G$.
Denoting by $L_{i} \equiv \gm\partial_{i}g$, $R_{i} \equiv \partial_{i}g\gm$
the left- and right-invariant vielbeins, one finds:
\beq
g_{ij} = - \tr(L_{i}L_{j}),\,\,\,\,
{\cal T}_{ij}^{k} = \frac{1}{2}
\tr(L_{k}(\partial_{i}L_{j}-\partial_{j}L_{i})).
\eeq
In the following we denote by $\Gamma_{ij}^{k}$ the Christoffel connection
associated to the metric $g_{ij}$.
Let  $\{T_{a}\}$ be a basis of ${\cal G}$, $L_{i} = L_{i}^{a} T_{a}$,  $R_{i} =
R_{i}^{a} T_{a}$. We may then define two sets of tangent space
coordinates for a vector ${\cal V}$:
\beq
V^{a}_{L} \equiv L_{i}^{a} {\cal V}^{i} \esp  V^{a}_{R} \equiv R_{i}^{a} {\cal
V}^{i}.
\eeq

The tangent space metric is $g_{ab} = -\tr(T_{a}T_{b})$, while the tangent
space components of the torsion  are proportional to the structure constants of
${\cal G}$.

	\subsection{Equations of motion and their solutions}

The classical equations of motion are
\beq
\partial_{--}(\gm\db{++}) =0 \Leftrightarrow \partial_{++}(\partial_{--} g \gm)
= 0. \label{eq1}
\eeq
Their general solution is a superposition of left- and right- movers:
\beq
g(x^{0},x^{1}) = \gl(-\xm)\gr(\xp).
\eeq
{}From the periodicity of $g$ we deduce that the monodromies of $\gl$ and $\gr$
defined as $\gamma_{L}(\xm) \equiv g^{-1}_{L}(-\xm) g_{L}(-\xm+2\pi)$ and
$\gamma_{R}(\xp) \equiv g_{R}(\xp) g^{-1}_{R}(\xp + 2 \pi)$ are equal and
therefore constant,
\beq
\gamma_{L}(\xm)=\gamma_{R}(\xp)=\gamma.
\eeq
We denote the left invariant currents by $j_{0,1} =g^{-1} \partial_{0,1} g $,
and
the chiral currents by $j_{L} \equiv \partial_{--} g g^{-1}$, $j_{R} \equiv
g^{-1} \partial_{++} g$.

\section{The chiral WZNW model}
\setcounter{equation}{0}
\label{eq105}
We follow here the approach of \cite{falc93,falc91,gawe91}.

	\subsection{Definition, phase space}

The phase space $P$ of the WZNW model may also be parameterized by the
functions $\gl$, $\gr$ and their monodromy $\gamma$. But these functions are
not uniquely fixed by a given solution $g(x,t)$ since $(\gl,\gr,\gamma)$ and
$(\gl g_{0}^{-1},g_{0}\gr,g_{0}\gamma g_{0}^{-1})$ for constant elements $g_{0}
\in G$ give the same solution. Introducing the phase spaces $P_{L}$ $(P_{R})$
of left- (right-)movers as the spaces of smooth maps $g_{L}$ $(g_{R})$ with
monodromies $\gamma_{L}$ ($\gamma_{R}$), and denoting by $\Delta$ the subset of
$P_{L} \times P_{R}$ with equal monodromies $\ml = \mr$, we have $P = \Delta/G$
where $G$ acts by
\beq
(g_{L},g_{R},\gamma) \rightarrow (g_{L} g^{-1}_{0},  g_{0} g_{R}, g_{0}\gamma
g_{0}^{-1}). \label{eq70}
\eeq

In the following we shall consider the left- and right-movers separately and
regard the phase space $P$ as a symplectic quotient of a larger space $P_{L}
\times P_{R}$ in which we relax the constraint that the monodromy of $g_{L}$
and $g_{R}$ be the same.
The symplectic 2-form of the WZNW model may be divided into two terms,
$\Omega_{L}$ and  $\Omega_{R}$. In this process there appear a  $2$-form $\rho$
on $G$ which has to be chosen such that
\beq
d\Omega_{R} = - (12 \pi)^{-1} \tr (d\mr \mr^{-1})^{3} - (4 \pi)^{-1}
d\rho(\mr)=0.
\eeq

If $r_c$ is a classical $r$-matrix on ${\cal G}$, then a possible choice for
the 2-form $\rho$ is given in terms of the decomposition $\gamma =
(\gamma^{-})^{-1} \gamma^{+}$ induced by $r_c$ as \cite{falc93,furl95}:
$\rho(\gamma) \equiv \tr(d \gamma^{-} ({\gamma^{-}})^{-1} \wedge d \gamma^{+}
({\gamma^{+}})^{-1} )$. Such a choice leads to the following
Poisson bracket on $P_{R}$:
\bea
&\pcc{\gr}{\gr} = \frac{1}{2} ({r_{c}}_{12} + C_{12} \, \epz )
{g_{R}}_{1}(x){g_{R}}_{2}(y) \label{eq23} \\
&\{ {g_{R}}_{1}(x), {\gamma_{R}}_{2} \} = (\rc^{+} {\mr}_{2} - {\mr}_{2}
\rc^{-}) {g_{R}}_{1}(x) \label{eq76} \\
&\{ {\gamma_{R}}_{1}, {\gamma_{R}}_{2} \} = \rc^{+} {\mr}_{1} {\mr}_{2} -
{\mr}_{1} \rc^{+} {\mr}_{2} - {\mr}_{2} \rc^{-} {\mr}_{1} + {\mr}_{1} {\mr}_{2}
\rc^{-} \label{eq77}
\eea
where $C$ is the Casimir operator, $|x-y| < 2 \pi$ and $\varepsilon^{0}$ is the
sign distribution defined in the appendix. $P_{R}$ equipped with this Poisson
bracket is known as the {\em chiral} WZNW model \cite{fadd90, alek90,
blok89,babe88,chu91}.

It is useful to recall how the Jacobi identity is satisfied by the bracket
(\ref{eq23}), and later compare with the $N=2$ case.
The sum $\{ \{ {\gr}_{1}(x), {\gr}_{2}(y)\}, {\gr}_{3}(z) \} + c.p.$, takes the
form
\beq
\frac{1}{4} A_{123}(x,y,z) {\gr}_{1}(x) {\gr}_{2}(y) {\gr}_{3}(z) + c.p.
\eeq
where $A$ splits into three parts. The terms independent of $r_c$ are
\beq
\varepsilon^{0}_{xy} C_{12}(\varepsilon^{0}_{xz} C_{13} + \varepsilon^{0}_{yz}
C_{23}) + c.p. = (\varepsilon^{0}_{xy} \varepsilon^{0}_{xz} + c.p. ) [ C_{12} ,
C_{13} ]. \label{eq37}
\eeq
The terms linear in $\rc$ automatically vanish, whereas the terms quadratic in
$\rc$ give the left hand side of the classical modified Yang-Baxter equation:
$\left[\rc_{12},\rc_{13}\right]+\left[ \rc_{12},\rc_{23} \right]+
\left[\rc_{13},\rc_{23} \right] = - [ C_{12} , C_{13} ]$. Finally one obtains
\beq
A_{123}(x,y,z) = (\varepsilon^{0}_{xy} \varepsilon^{0}_{xz} +
\varepsilon^{0}_{yz} \varepsilon^{0}_{yx} + \varepsilon^{0}_{zx}
\varepsilon^{0}_{zy} -1 ) [ C_{12} , C_{13} ] = 0 \label{eq95}
\eeq
where we have used the fundamental property (\ref{eq25}) given in the appendix
of the sign distribution $\varepsilon^{0}$.

	\subsection{Symmetries of the chiral WZNW model}

The chiral WZNW model possesses a rich symmetry structure.

The conformal symmetry is parametrized by monotonous maps $f: \mbox{\bf R}
\rightarrow \mbox{\bf R}$ acting on the chiral field  $g_{R}$ as: $g_{R}
\rightarrow g_{R} \circ f$. It is generated at the infinitesimal level by the
energy momentum tensor $T =-\frac{1}{2} \tr j_{R}^{2}$. The Hamiltonian of the
chiral WZNW model is given by
$H = \int dx T$.
It leads to the equation of motion
\beq
\partial_{t} g_{R} = \{ g_{R}, H \} \Leftrightarrow\partial_{--} g_{R} = 0,
\eeq

The model is also invariant under a loop group symmetry. The infinitesimal
action of the loop group $LG$ is given by $\delta g_{R} = g_{R}
\omega(x^{++})$, where $\omega$ is a $2\pi$-periodic function taking its values
in the Lie algebra of $G$. It is generated by the currents $j_{R}$. Finally,
there is also a Poisson-Lie symmetry. It is given by  the left action of the
Poisson-Lie group $G_{PL}$ equipped with the Sklyanin bracket, i.e.
\beq
G_{PL} \times P_{R} \ni (h,g_{R}) \rightarrow h g_{R} \in P_{R}
\eeq
where
\beq
\{ h_{1}, h_{2} \} = \left[ \rc^{\pm}_{12}, h_{1}h_{2} \right]. \label{eq96}
\eeq

The Hamiltonian is invariant under this action, and the Poisson brackets
(\ref{eq23}) transform covariantly.
The moment map is given by \cite{falc93}
\beq
g_{R} \rightarrow (\gamma_{R}^{+}, \gamma_{R}^{-}) \in G_{PL}^{*}
\eeq
where $\gamma_{R} = (\gamma_{R}^{-})^{-1} \gamma_{R}^{+}$. The Poisson brackets
of $\gamma_{R}^{+}$ with the phase space variables $g_R(y)$ is given by
\bea
\{ {\gamma_{R}}_{1}^{+}, {g_{R}}_{2}(y) \} &=& - {\gamma_{R}}_{1}^{+}
\rc_{12}^{+} g_{2}(y), \\
\{ {\gamma_{R}}_{1}^{-}, {g_{R}}_{2}(y) \} &=& - {\gamma_{R}}_{1}^{-}
\rc_{12}^{-} g_{2}(y),
\eea
so that the infinitesimal action of $G_{PL}$ is
\beq
\tr_{1} (\lambda_{1} ({\gamma_{R}}_{1}^{+})^{-1} \{ {\gamma_{R}}_{1}^{+},
{g_{R}}_{2}(y) \}) = \rc^{-} (\lambda) g_{R}.
\eeq

\section{The $N=1$ WZNW model}
\setcounter{equation}{0}
\label{eq100}
The goal of this section is to review the construction and the properties of
the $N=1$ WZNW model: within the framework of $N=1$ superfields and superspace,
one can construct an explicitly supersymmetric invariant action. Then one
immediately finds that the properties of the $N=1$ model are similar to the
ones of the bosonic model.

	\subsection{$N=1$ superspace}

The generators of the $N=1$ supersymmetry algebra will be denoted by
$Q_{+,-}^{1}$,
\beq
\{Q_{+}^{1},Q_{+}^{1}\} = -2 \partial_{++},\,\,
\{Q_{-}^{1},Q_{-}^{1}\} = -2 \partial_{--},\,\,
\{Q_{+}^{1},Q_{-}^{1}\} = 0.\eeq
The coordinates of the $N=1$ superspace $\cal M$ are
$(x^{++},x^{--},\theta^{+}_{1},\theta^{-}_{1})$. The supersymmetry generators
$Q_{\pm}^{1}$ may then be realized as
\beq
Q_{\pm}^{1} = \frac{\partial}{\partial{\theta^{\pm}_{1}}} -
\theta^{\pm}_{1}\partial_{\pm\pm}.
\eeq
The spinor covariant derivatives $D_{+,-}^{1}$, anticommuting with the
supersymmetry generators, are
\beq
D_{\pm}^{1} = \frac{\partial}{\partial{\theta^{\pm}_{1}}} +
\theta^{\pm}_{1}\partial_{\pm\pm}.
\eeq
The $(1,1)$ superfields $\phi^{i}(x,\theta^{+}_{1},\theta^{-}_{1})$ have as
lowest component the ordinary field $\phi^{i}(x,0,0) = \varphi^{i}(x)$. The
$(1,1)$ supersymmetry transformations are
\beq
\delta_{\epsilon}^{1}\phi^{i} \equiv
\epsilon^{+}Q_{+}^{1}\phi^{i}+\epsilon^{-}Q_{-}^{1}\phi^{i}.
\eeq
The manifestly $N=1$ invariant action which may be seen as a
supersymmetrization of (\ref{eq71}) is obtained by replacing the integration on
$M$ by an integration on the superspace ${\cal M}$, the fields $\varphi^{i}$ by
the superfields $\phi^{i}$ and the vector derivatives by covariant spinor
derivatives:
\beq
S[\phi] \equiv \int_{\cal M} dx^{++}dx^{--}d\theta^{+}_{1}d\theta^{-}_{1}
(g_{ij} + b_{ij})(\phi)D_{+}^{1}\phi^{i}D_{-}^{1}\phi^{j}. \label{eq3}
\eeq
In the case of the WZNW nodel, this may be written in terms of the
group valued superfield $g: {\cal M} \rightarrow G$, which we will denote by
the same letter as the ordinary field in (\ref{eq2}),
\bea
S[g] &=& \int_{\cal M} d^{2}x  d^{2} \theta_{1} \tr[ (g^{-1} D_{+}^{1}g)(g^{-1}
D_{-}^{1}g) \nonumber \\
&+& \int dt  (g^{-1}\partial_{t}g)((g^{-1}D_{+}^{1}g)(g^{-1}D_{-}^{1}g) +
(g^{-1}D_{-}^{1}g)(g^{-1}D_{+}^{1}g))].	\label{eq72}
\eea
Notice that the superfield $g$ still belongs to the group $G$. We will not
consider here the case of a supergroup.
{}From the action (\ref{eq72}) we derive the equations of motion
\beq
D_{-}^{1}(g^{-1} D_{+}^{1}g) = 0 \Leftrightarrow D_{+}^{1}(D_{-}^{1}g g^{-1}) =
0.
\eeq
The general solution is still given by a superposition of left- and
right-movers $g(x^{\pm\pm},\theta^{\pm}_{1}) = g_{L} (-x^{--},\theta^{-}_{1})
g_{R} (x^{++},\theta^{+}_{1})$ of constant and equal monodromies (and therefore
equal to the monodromies of the underlying bosonic WZNW model).
The supercurrents $j_{L} \equiv D_{-}^{1}g g^{-1}$ and $j_{R} \equiv
g^{-1}D_{+}^{1}g$ have the following Poisson brackets (see the appendix for
notations)
\bea
\{ {j_{L}}_1(\mathbf{x}) , {j_{L}}_2(\mathbf{y}) \} &=&
-\left[C_{12},{j_{L}}_{1}(\mathbf{x})\right]\dexyu - C_{12}
{D_{+}^{1}}_{\mathbf{x}} \dexyu,\\
\{ {j_{R}}_1(\mathbf{x}) , {j_{R}}_2(\mathbf{y}) \} &=&
-\left[C_{12},{j_{R}}_{1}(\mathbf{x})\right]\dexyu + C_{12}
{D_{+}^{1}}_{\mathbf{x}} \dexyu,\\
\{ {j_{L}}_1(\mathbf{x}) , {j_{R}}_2(\mathbf{y}) \} &=& 0
\eea
which means that the phase space of the $N=1$ WZNW model contains two commuting
copies of an $N=1$ Kac-Moody algebra.

	\subsection{The $N=1$ chiral WZNW model and its symmetries}

Exactly the same steps as in the bosonic case lead to the chiral Poisson
brackets
\bea
& \{ {g_R}_1(\mathbf{x}) , {g_R}_2(\mathbf{y}) \} = \frac{1}{2} ({r_{c}}_{12} +
C_{12} \, \epu) {g_{R}}_{1}(\mathbf{x}){g_{R}}_{2}(\mathbf{y}), \label{eq24} \\
&\{ {g_{R}}_{1}(\mathbf{x}), {\gamma_{R}}_{2} \} = (\rc^{+} {\mr}_{2} -
{\mr}_{2} \rc^{-}) {g_{R}}_{1}(\mathbf{x}), \\
&\{ {\gamma_{R}}_{1}, {\gamma_{R}}_{2} \} = \rc^{+} {\mr}_{1} {\mr}_{2} -
{\mr}_{1} \rc^{+} {\mr}_{2} - {\mr}_{2} \rc^{-} {\mr}_{1} + {\mr}_{1} {\mr}_{2}
\rc^{-}. \label{eq93}
\eea
where $|x-y|< 2 \pi$ and $\varepsilon^{1}$ is the $(1,0)$ sign distribution
defined in the appendix.
Again the $N=1$ chiral WZNW model possesses a rich symmetry structure which we
briefly review. The $N=1$ superconformal symmetry acts at the infinitesimal
level by $\delta g_{R} \equiv \epsilon \partial g_{R} + \frac{1}{2} D \epsilon
Dg_{R}$. The corresponding moment map is given by the superconformal energy
momentum tensor
\beq
T \equiv \frac{1}{3} \tr (j_{R}^{3}) + \frac{1}{2} \tr (j_{R} Dj_{R})
\eeq
and the Hamiltonian is $H \equiv \int dx d\theta_{1}^{+}  T$.
The $N=1$ loop group symmetry corresponds to right translations generated by
the current $j_{R}$. Finally, the Poisson-Lie symmetry corresponding to the
action of $G_{PL}$ is the same as in the bosonic case. It is still generated by
the monodromy.

In conclusion for {\em any} group $G$ we can always construct an $N=1$
extension of the bosonic WZNW model. We are now in position to consider the
$N=2$ extension.
\section{The $N=2$ WZNW model}
\setcounter{equation}{0}
\label{eq101}
	\subsection{Conditions on the geometry}

We now want to make the action (\ref{eq3}) invariant under two additional
supersymmetries forming with the previous ones an $N=2$ supersymmetry algebra
with $SO(1,1)$ automorphism
group:
\beq
\{Q_{\pm}^{1},Q_{\pm}^{1}\} =  -2 \partial_{\pm\pm}, \,\,
\{Q_{\pm}^{2},Q_{\pm}^{2}\} =  +2 \partial_{\pm\pm}, \label{eq60} \,\,
\{Q_{\pm}^{1},Q_{\pm}^{2}\} =  0.
\eeq
In contrast to the $N=1$ case, the existence of an extended supersymmetry
requires some conditions on the geometry of the target manifold. These
conditions have been studied in \cite{alva81} for a target space without
torsion and in \cite{gates,howe,hull85,hull87} for a target space with torsion.
The particular case of a  group manifold was studied in
\cite{sevr88-1,sevr88-2,sevr88-3}. Let us recall how these conditions come out.
The most general form of a second supersymmetry transformation commuting with
the first one is:
\beq
\delta_{\epsilon}^{2}\phi^{i} \equiv \epsilon^{+}{\cal
R}^{i}_{\hspace{4pt}j}(\phi)D_{+}^{1}\phi^{j}+\epsilon^{-}{\tilde {\cal
R}}^{i}_{\hspace{4pt}j}(\phi)D_{-}^{1}\phi^{j} \label{eq4}
\eeq
where ${\cal R}$ and ${\tilde {\cal R}}$ are tensors.
The invariance of the action (\ref{eq3}) under (\ref{eq4}) implies first

\beq
g_{ik}{\cal R}^{k}_{\hspace{4pt}j} = - g_{kj}{\cal R}^{k}_{\hspace{4pt}i} \esp
g_{ik}{\tilde {\cal R}}^{k}_{\hspace{4pt}j} = - g_{kj}{\tilde {\cal
R}}^{k}_{\hspace{4pt}i}. \label{eq6}
\eeq
Second, ${\cal R}$ and ${\tilde {\cal R}}$ have to be covariantly constant with
respect to different connections:
\beq
{\cal D}_{i}^{-}{\cal R}^{j}_{\hspace{4pt}k} = 0 \esp {\cal D}^{+}_{i}{\tilde
{\cal R}}^{j}_{\hspace{4pt}k} = 0 \label{eq7}
\eeq
where ${\cal D}^{\pm}_{i}{\cal V}_{j} \equiv \partial_{i}{\cal V}_{j} +
(\Gamma_{ij}^{k} \pm {\cal T}_{ij}^{k}){\cal V}_{k}$.
Next, requiring the supersymmetry algebra (\ref{eq60}) to be satisfied by the
new transformations (\ref{eq4}) leads to
\beq
{\cal R}^{2} = 1 \qquad \mbox{and} \qquad {\tilde {\cal R}}^{2} = 1,
\label{eq5}
\eeq
and to the vanishing of the Nijenhuis tensor of ${\cal R}$ and ${\tilde {\cal
R}}$
\beq
N^{i}_{\hspace{4pt}jk}({\cal R}) =   {\cal R}^{l}_{\hspace{4pt}[k} {\cal
R}^{i}_{\hspace{4pt}j],l}  -
{\cal R}^{l}_{\hspace{4pt}[j,k]} {\cal R}^{i}_{\hspace{4pt}l} =0,\,\,
N^{i}_{\hspace{4pt}jk}(\tilde{\cal R})=0. \label{eq8}
\eeq
If we further require the supersymmetry algebra to close without any use of the
equations of motion, we find one more condition:
\beq
\left[{\cal R},{\tilde {\cal R}}\right] = 0.
 \label{eq14}\eeq

	\subsection{The Nijenhuis tensor and the modified Yang-Baxter equation}

The conditions just written and their solutions were studied in great detail,
in the case of a supersymmetry algebra with $SO(2)$ automorphism group,
in \cite{sevr88-1,sevr88-2,sevr88-3}, which we now follow.
The first step is to use the covariant constancy of ${\cal R}$ together with
the vanishing of the Nijenhuis tensor
 to find an equation involving ${\cal R}$  and the torsion coefficients:
\beq
{\cal T}_{jk}^{i} + {\cal T}_{lm}^{i} {\cal R}^{l}_{\hspace{4pt}j}{\cal
R}^{m}_{\hspace{4pt}k} - {\cal T}_{jl}^{m}{\cal R}^{i}_{\hspace{4pt}m}{\cal
R}^{l}_{\hspace{4pt}k} - {\cal T}_{lk}^{m} {\cal R}^{i}_{\hspace{4pt}m}{\cal
R}^{l}_{\hspace{4pt}j}  =0. \label{eq59}
\eeq
There is an identical equation involving  ${\tilde {\cal R}}$.
It is easier to carry out the discussion  in the tangent space,  which here is
the Lie algebra ${\cal G}$. We define
\beq
r^{a}_{\hspace{4pt}b} = L_{i}^{a} L_{b}^{j} {\cal R}^{i}_{\hspace{4pt}j},
\,\,\,\,
{\tilde r}^{a}_{\hspace{4pt}b} =  R_{i}^{a} R_{b}^{j} {\tilde {\cal
R}}^{i}_{\hspace{4pt}j},
\label{bof1}\eeq
where we have used the notations of  section \ref{eq106}.

In tangent space the conditions (\ref{eq6}), (\ref{eq7}), (\ref{eq5}) and
(\ref{eq59}) become respectively:
\bea
&g_{ab}r^{b}_{\hspace{4pt}c} = - g_{bc}r^{b}_{\hspace{4pt}a},\,\,
\partial_{i} r^{a}_{\hspace{4pt}b} = 0, \,\,
r^{a}_{\hspace{4pt}b}r^{b}_{\hspace{4pt}c} = 1 ,&\\
&f_{bc}^{\hspace{8pt}a} + f_{de}^{\hspace{8pt}a} r^{d}_{\hspace{4pt}b}
r^{e}_{\hspace{4pt}c} - f_{bd}^{\hspace{8pt}e} r^{a}_{\hspace{4pt}e}
r^{d}_{\hspace{4pt}c}  -  f_{dc}^{\hspace{8pt}e} r^{a}_{\hspace{4pt}e}
r^{d}_{\hspace{4pt}b} = 0.&
\eea
Equivalently, we may define a linear operator $r$ on ${\cal G}$  by $r(T_{a})
\equiv r^{b}_{\hspace{4pt}a} T_{b}$. Then, for any $M$ and $N$ in ${\cal G}$
we have
\bea
& \tr r(M)N = - \tr \, Mr(N),\,\,
\partial_{i} r = 0,\,\,
r^{2}(M) = M,
  \label{eq9} \\
&\left[r(M),r(N)\right] - r\left[M,r(N)\right] - r\left[r(M),N\right]
=-\left[M,N\right].&\label{eq22}
\eea
There are identical equations for ${\tilde r}$.
The last condition is the modified Yang-Baxter equation.
The matrix $r$ is thus a classical antisymmetric $r$-matrix, with the extra
requirement that the square of $r$ is the identity.
Finally if we require that the supersymmetry algebra closes without any use of
the equations of motion, condition (\ref{eq14}) becomes
\beq
\left[r,Ad(g){\tilde r}Ad(g^{-1})\right] =0, \forall g\in G. \label{eq20}
\eeq

Since the square of $r$ is the identity, its eigenvalues are $\pm 1$,
and we denote by $\allie_{\pm}$ the two eigenspaces.
Then conditions (\ref{eq9},\ref{eq22}) mean that both $\allie_{+}$
and $\allie_{-}$ are isotropic subalgebras of the Lie algebra $\cal G$.
A subalgebra is isotropic if the restriction of the invariant
quadratic form to it vanishes. The data $(\allie,\allie_{+},\allie_{-})$
is called a Manin triple. This interpretation of the constraints
associated with $N=2$ supersymmetry was given and used by
Parkhomenko \cite{park92} and further by Getzler \cite{getz93-2}.

When $\allie$ is a complex reductive Lie algebra, the results obtained in
\cite{sevr88-1} may be applied to the construction of triples.
The only condition for the existence of solutions is that the Lie
algebra $\allie$ should be even dimensional.
The construction of triples goes as follows. First choose
 a Cartan decomposition of $\allie$, $\allie = {\cal N}_{-} \oplus {\cal H}
\oplus {\cal N}_{+}$. Then decompose the Cartan subalgebra ${\cal H}$ as
${\cal H} = {\cal H}_{+} \oplus {\cal H}_{-}$, with $\mbox{\rm dim}{\cal H}_{+}
=\mbox{\rm dim}{\cal H}_{-}$, and ${\cal H}_{\pm}$ isotropic. Then one may take
$\allie_{\pm} \equiv {\cal N}_{\pm} \oplus {\cal H}_{\pm}$, and any triple may
be obtained in this way. The r-matrix associated to a given
triple generically do not act on the real forms of the
complex Lie algebra $\allie$. However, if one may find a basis of
$\allie_+$ and a basis of $\allie_-$ such that all structure constants are
real, then the real linear combinations of these basis elements
form a real algebra on which $r$ acts. In other words, this real form does
admit a Manin triple. This possibility will be realized on non-simple
real algebras which are sum of  two (not necessarily simple) algebras of
equal ranks.
Finally, it was proved in \cite{roce91-2} that the only complex Lie groups
admitting two solutions of the system (\ref{eq9},\ref{eq22}) and verifying
(\ref{eq20}) are $(SL(2) \otimes GL(1))^{n} \otimes GL(1)^{2q}$.
Only for these groups, or for real forms of this group, does the supersymmetry
algebra close without the use of the equations of motion. Since it is so
restrictive, we shall not require in the following constraint (\ref{eq20})
to hold.

\section{The $N=2$ WZNW in $(2,2)$ superspace}
\setcounter{equation}{0}
\label{eq102}
The goal of the next sections is to exhibit the properties of the $N=2$ WZNW
model such as equations of motion, constraints, symmetries, Poisson brackets.
This work will be carried out in the $(2,2)$ superspace in order to make
explicit the $N=2$ invariance of the model.

	\subsection{Equations and constraints}

We have to write all the equations defining the model in terms of $(2,2)$
superfields.
More precisely, we add anti-commuting coordinates
$(\theta^{+}_{2},\theta^{-}_{2})$ which form with
$(x^{\pm\pm},\theta^{\pm}_{1})$ the $(2,2)$ superspace. Then we define
$Q_{\pm}^{2}$ and $D_{\pm}^{2}$ by
\beq
Q_{\pm}^{2} = \frac{\partial}{\partial{\theta^{\pm}_{2}}} +
\theta^{\pm}_{2}\partial_{\pm\pm},\,\,
D_{\pm}^{2} = \frac{\partial}{\partial{\theta^{\pm}_{2}}} -
\theta^{\pm}_{2}\partial_{\pm\pm}.
\eeq
The $(1,1)$ and $(2,2)$ superfields will be respectively denoted by lowercase
and
uppercase letters such that
$C(x,\theta^{\pm}_{1},\theta^{\pm}_{2}) |_{\theta^{\pm}_{2} =
0}=c(x,\theta^{\pm}_{1}).$
In the $(1,1)$ superspace the equations of motion are
$D_{+}^{1}(D_{-}^{1}gg^{-1}) = 0$. This requires the (2,2) superfield $G$ to
satisfy the equation of motion
\beq
D_{+}^{1}(D_{-}^{1}GG^{-1}) = 0 \label{eq11},
\eeq
such that we recover the previous equation of motion when $\theta^{\pm}_{2} =
0$. In superspace, the second supersymmetry transformations are realized with
the help of the derivatives $Q^2_\pm$.
The transformation of the $(2,2)$ superfield $\Phi^{i}$ gives, when $\theta_{2}
=0$, the transformation of its $N=1$ component $\phi^{i}$,  for which we
already have an expression in equation \reff{eq4}. The equality of these two
expressions leads to constraints on the $N=2$ superfields.
More precisely, if we consider for the time being only the $\epsilon^+$
transformation, $\Phi^{i}$ transforms as $\delta_{\epsilon}^{2}\Phi^{i} \equiv
\epsilon^{+}Q_{+}^{2}\Phi^{i}$ which gives
\beq
\delta_{\epsilon}^{2}\phi^{i} =
(\delta_{\epsilon}^{2}\Phi^{i})_{\theta^{\pm}_{2} = 0} =
\epsilon^{+}(Q_{+}^{2}\Phi^{i})_{\theta^{\pm}_{2} = 0}
=\epsilon^{+} (D_{+}^{2}\Phi^{i})_{\theta^{\pm}_{2} = 0}.
\label{eq12}\eeq
But we have already found in eq.(\ref{eq4}) the expression \beq
\delta_{\epsilon}^{2}\phi^{i} \equiv \epsilon^{+}{\cal
R}^{i}_{\hspace{4pt}j}D_{+}^{1}\phi^{j}
= \epsilon^{+} ({\cal
R}^{i}_{\hspace{4pt}j}D_{+}^{1}\Phi^{j})_{\theta^{\pm}_{2}=0} \label{eq13}
\eeq
Equating the right-hand sides of eqs.(\ref{eq12}) and (\ref{eq13}) leads to the
constraints
\bea
(D_{+}^{2}\Phi^{i})_{\theta^{\pm}_{2} = 0} &=& ({\cal
R}^{i}_{\hspace{4pt}j}D_{+}^{1}\Phi^{j})_{\theta^{\pm}_{2} = 0} \\
(D_{-}^{2}\Phi^{i})_{\theta^{\pm}_{2} = 0} &=&  ({\tilde {\cal
R}}^{i}_{\hspace{4pt}j}D_{-}^{1}\Phi^{j})_{\theta^{\pm}_{2} = 0}
\eea
which we have to extend to the whole superspace for consistency with $N=2$
supersymmetry.
Using equation \reff{bof1}, we then find that the
group valued superfield $G$ satisfies the constraints
\bea
G^{-1} D_{+}^{2}G &=& r(G^{-1}D_{+}^{1}G), \label{eq18} \\
D_{-}^{2}G G^{-1} &=& {\tilde r}(D_{-}^{1}G G^{-1}) \label{eq17}
\eea
together with the equation of motion (\ref{eq11}). In fact, taking into account
the constraints, we may write the equations of motion in the more symmetric
form
\beq
D_{+}^{i}(D_{-}^{j}GG^{-1}) = 0 \Leftrightarrow D_{-}^{i}(G^{-1}D_{+}^{j}G) =0
,\,\, i,j=1,2. \label{eq50}
\eeq
	\subsection{Constraints and dynamics}

The next step for the construction of a theory explicitly invariant under $N=2$
supersymmetry would be the construction of an action involving $(2,2)$
superfields, as it was done in the $N=1$ case. This is a difficult task since,
as we will now show, the constraints (\ref{eq18}) and (\ref{eq17}) imply the
equation of motion (\ref{eq50}).
Let us start from the identity
\beq
D_{-}^{i}(G^{-1}D_{+}^{j}G) = - G^{-1}D_{+}^{j}(D_{-}^{i}G G^{-1})G
=-Ad(G)D_{+}^{j}(D_{-}^{i}G G^{-1}) \label{eq16}
\eeq
which relates one constraint to the other.
Starting from the constraint (\ref{eq18}) we have
\beq
D_{-}^{1}(G^{-1}D_{+}^{1}G) = r(D_{-}^{1}(G^{-1}D_{+}^{2}G)).
\eeq
Then, using the relation (\ref{eq16}) with $i=1,j=2$ and the constraint
(\ref{eq17}), we find
\beq
D_{-}^{1}(G^{-1}D_{+}^{1}G ) = -r \{ Ad(G){\tilde r} \left[ D_{+}^{2} (
D_{-}^{2}GG^{-1} ) \right]\}.
\eeq
Using once again the relation (\ref{eq16}) with $i=j=2$ leads to
\beq
D_{-}^{1}(G^{-1}D_{+}^{1}G ) = r Ad(G) {\tilde r}Ad(G^{-1}) \{ D_{-}^{2}
(G^{-1} D_{+}^{2} G) \}.
\eeq
Repeating exactly the same operations will again permute $1$ and $2$ and give
\beq
D_{-}^{1}(G^{-1}D_{+}^{1}G ) = ( r Ad(G) {\tilde r}Ad(G^{-1}))^{2}
 \{ D_{-}^{1} (G^{-1} D_{+}^{1} G) \}
\eeq
which may be written as
\beq
\left[ r ,Ad(G) {\tilde r}Ad(G^{-1})\right] (D_{-}^{1}(G^{-1}D_{+}^{1}G)) =0.
\eeq

Thus, the condition for the constraints (\ref{eq17},\ref{eq18})
not to imply the equation of motion is the same as the condition (\ref{eq20})
found in $N=1$ superspace for the extended supersymmetry algebra to close
without the use of the equations of motion. It is only in this case
that an explicitly supersymmetric action is known. Such an action was
constructed for the case of $SU(2) \otimes U(1)$ in \cite{ivan,roce91-2}.
In all other cases, the constraints imply at least part of the equations of
motion, and no explicitly invariant action is known. In the next section
we shall study the properties of the model using the constraints and the
equation of motion.

	\subsection{Chiral derivatives}

{}From now on we shall make the simplifying choice $r \equiv {\tilde r} \equiv
\rmt$.
Let us define $\rpmt \equiv \frac{1}{2} (\rmt \pm Id)$. For any element $a \in
{\cal G}$ we write $a^{(\pm)} = r^{\pm}_{m}(a)$, so that $a = a^{(+)} -
a^{(-)}$.
We define $\theta^{+} \equiv \theta_{1}^{+} + \theta_{2}^{+}$, ${\bar
\theta}^{+} \equiv \theta_{1}^{+} - \theta_{2}^{+}$ and the chiral $D_{\pm}$
and anti-chiral $\bar D_{\pm}$ derivatives:
\beq
D_{\pm} \equiv \frac{1}{2} ( D_{\pm}^{1} +  D_{\pm}^{2}) \esp {\bar D}_{\pm}
\equiv \frac{1}{2} (D_{\pm}^{1} - D_{\pm}^{2}).
\eeq
They satisfy $(D_{\pm})^{2} =0$, $({\bar D}_{\pm})^{2} =0$ and $\{
D_{\pm},{\bar D}_{\pm} \} = \partial_{\pm\pm}$.
The corresponding supercurrents are
\beq
J_+ = G^{-1} D_{+}G, \,\,\,
{\bar J_+} = G^{-1} {\bar D}_{+}G.
\label{eqq1}\eeq
{}From this definition (\ref{eqq1}) of the chiral supercurrents and
from the properties of the chiral spinor derivatives we deduce that
$J_+$ and ${\bar J_+}$ satisfy the non-linear constraints
\beq D_{+}J_++J_+J_+=0,\,\,\,
{\bar D}_{+}{\bar J_+}+{\bar J_+}{\bar J_+}=0.
\label{eqq2}\eeq
Moreover, from eq.(\ref{eq18}-\ref{eq17}) we find that
$J_+$ and ${\bar J_+}$ satisfy the linear constraints
\beq
J_{+}^{(-)} =0 \esp {\bar J}_{+}^{(+)} = 0. \label{eq44}
\eeq
The property that $\allie_+$ and $\allie_-$ are algebras ensures the
consistency of these two sets of constraints. Finally, the equations of motion
may be written as
\beq
D_{-}J_+ = {\bar D}_{-}J_+ = D_{-} {\bar J}_+ = {\bar D}_{-} {\bar J}_+ = 0.
\label{eq79}
\eeq

	\subsection{Solutions of the equations of motion}

As in the bosonic and $N=1$ cases, the general solution of the equation of
motion is a superposition of left- and right-movers with constant and equal
monodromies:
\beq
G(x^{\pm},\theta_{1}^{\pm}, \theta_{2}^{\pm}) = G_{L}(-x^{--},\theta_{1,2}^{-})
G_{R}(x^{++},\theta_{1,2}^{+})
\eeq
constrained by
\bea
r_{m}^{-} (G_{R}^{-1} D_{+} G_{R}) =0 &\esp& r_{m}^{+} (G_{R}^{-1} {\bar D}_{+}
G_{R}) = 0, \label{eq92} \\
r_{m}^{-} ( D_{-} G_{L} G_{L}^{-1} ) = 0 &\esp& r_{m}^{+} ({\bar D}_{-} G_{L}
G_{L}^{-1}) =0.
\eea
 It is natural in order to solve the constraints (\ref{eq92}) on $G_{R}$ to use
the decomposition $G_{R} = G_{R}^{(-)} G_{R}^{(+)}$ induced by $r_{m}$. However
if the first one gives $D_+ G_{R}^{(-)} = 0$, the second one is much more
complicated to solve. Therefore, it may be interesting to decompose also
$G_{R}$ in the opposite way, $G_{R} = {\tilde G}_{R}^{(+)} {\tilde
G}_{R}^{(-)}$. Then the constraints give
\beq
D_+ G_{R}^{(-)} = 0 \esp {\bar D}_+ {\tilde G}_{R}^{(+)} = 0
\eeq
while $G_{R}^{(+)}$ and ${\tilde G}_{R}^{(-)}$ are determined in terms of
$G_{R}^{(-)}$ and ${\tilde G}_{R}^{(+)}$ through the factorization problem
\beq
({\tilde G}_{R}^{(+)})^{-1} G_{R}^{(-)} = {\tilde G}_{R}^{(-)}
(G_{R}^{(+)})^{-1}.
\eeq

	\subsection{Loop group transformations}

We now wish to study the symmetries of the model, starting with the loop group
symmetry already present in the $N=1$ case. It will be instructive to come back
to the $N=1$ superspace, and consider an infinitesimal global right
translation,
$ \delta_{\alpha} g = g \alpha$,
together with a second supersymmetry transformation,
\beq
\delta_{\epsilon}^2 g \equiv \epsilon^{+} g \rmt (j_{+}^{1}),\,\,
j^1_+=g^{-1}D^1_+g.
\eeq
The commutator of two such transformations reads
\beq
\left[ \delta_{\epsilon}^2, \delta_{\alpha} \right] g = \epsilon^{+} g \left(
\left[ r_{m}(j_{+}^{1}), \alpha \right] - r_{m} \left[ j_{+}^{1}, \alpha
\right] \right).
\eeq
Therefore, in general, the $N=1$ loop group symmetry does not commute with the
second supersymmetry. The only case where the commutator vanishes is if
$\alpha$ belongs to the Cartan subalgebra entering into the construction of the
matrix
$\rmt$ in section \ref{eq101}. A consequence is that in the $N=2$ superspace,
the parameter of the loop group symmetry, as we shall now see, is not an $N=2$
superfield.

More precisely, the generalization of the $N=1$ loop group symmetry in the
$(2,2)$ superspace is $\delta_{\Omega} G \equiv G \Omega$. Such a
transformation is a symmetry if it leaves the constraints (\ref{eq18}),
(\ref{eq17}) and  the equations of motion (\ref{eq11}) invariant. This leads to
\beq
\rmt^{-}(D_+ \Omega +[J_+,\Omega]) =0,\,\,
\rmt^{+}({\bar D}_+ \Omega
+[{\bar J}_+,\Omega]) =0,\label{eq61}
\eeq
together with $D_-\Omega={\bar D_-}\Omega =0$.
A convenient parametrization for the solutions of these equations is
\beq
\Omega = \omega + (\theta^{+} - {\bar \theta}^{+})( \left[\jb_{+},\omega
\right]^{(+)} + \left[J_{+},\omega\right]^{(-)} ), \label{eq62}
\eeq
with
\bea
D_{-} \omega &=& {\bar D}_{-} \omega =0,\\
D_{+} \omega^{(-)} &=& {\bar D}_{+} \omega^{(+)} =0.
\eea
One then finds that the commutator of two such transformations is given by
 $[\delta_{\Omega_{1}},\delta_{\Omega_{2}}] = \delta_{\Omega}$ with
\beq
\omega = - {\bar D}_{+}\{(\theta^{+} - {\bar \theta}^{+})
\left[\omega_{1},\omega_{2}\right]^{(+)}\} + D_{+} \{(\theta^{+} - {\bar
\theta}^{+}) \left[\omega_{1},\omega_{2}\right]^{(-)}\}.
\eeq
It is clear from eq.(\ref{eq62}) that the loop group parameters are not
superfields except, as we have already seen, when $\omega$ belongs to the
Cartan subalgebra ${\cal H}$, in which case $\Omega = \omega$.
We shall find in the next section  the currents generating these
transformations.

\section{The $N=2$ chiral WZNW model}
\label{eq103}
\setcounter{equation}{0}
We now study the Poisson brackets of the chiral $N=2$ WZNW model in superspace,
which is the main result of this article.
{}From it we shall recover the classical limit of the operator products
expansions of the currents found in \cite{hull90-2}.

	\subsection{Poisson brackets of the group elements}

The phase space of the chiral $N=2$ WZNW model is given by a superfield
$G(x,\theta_1,\theta_2)$ satisfying the constraints
\beq
r_{m}^{-} (G^{-1} D G) =0 \esp r_{m}^{+} (G^{-1} {\bar D} G) = 0, \label{bof2}
\eeq
and with constant monodromy $\gamma$. The expression of the Poisson brackets
may be found by expanding $G$ in components:
\beq
G(x,\theta_{1,2}) = g(x,\theta_{1})(1 + \theta_{2} \psi(x,\theta_{1})).
\eeq
such that $g = G|_{\theta_{2}=0}$ belongs to the phase space of a chiral $N=1$
WZNW model and carries the Poisson bracket (\ref{eq24}-\ref{eq93}).
The constraint $G^{-1} D^{2}G = \rmt(G^{-1}D^{1}G)$ at ${\theta_{2} = 0}$ then
determines $\psi$ in terms of $g$, $\psi = r_{m} (\gm D^{1} g)=r_{m}(j)$, so
that the $(2,2)$ superfield $G$ only depends on the $(1,1)$ superfield $g$:
\beq
G = g (1 + \theta_{2} r_{m}(j)).
\eeq
We can then compute the Poisson bracket of $G$ from the Poisson brackets of $g$
and $j$. This is a long computation because we wish to express the result in
terms of  $G$ alone. For this we need to use the modified Yang-Baxter equation
for $r_{m}$.
We also compute the Poisson bracket of $G$ with its monodromy, which is
immediate since $j$ commutes with $\gamma$. The results are the following:
\bea
&\{ G_{1X}, G_{2Y} \} =  \frac{1}{2} G_{1X} G_{2Y} {r_{m}}_{12}\dexyd
+ \frac{1}{2} ({r_{c}}_{12}+ C_{12} \varepsilon^{2}_{XY}) G_{1X} G_{2Y},
\label{eq31} \\
&\{ G_{1X}, \gamma_{2} \} = (\rc^{+} \gamma_{2} - \gamma_{2} \rc^{-}) G_{1X},
\\
&\{ \gamma_{1}, \gamma_{2} \} = \rc^{+} \gamma_{1} \gamma_{2} - \gamma_{1}
\rc^{+} \gamma_{2} - \gamma_{2} \rc^{-} \gamma_{1} + \gamma_{1} \gamma_{2}
\rc^{-}
\eea
where the Dirac and sign distributions $\delta^{2}$ and $\varepsilon^{2}$ are
defined in the appendix. Let us comment on this Poisson bracket.
First we may verify that in the $N=1$ limit, we recover the Poisson bracket of
$g$ since the first term in the
right-hand side of eq.(\ref{eq31}) vanishes.
Then we note that two different $r$-matrices appear. The first one $r_{c}$ is
already present at the bosonic level, while the second one $r_{m}$ only comes
from the $N=2$ constraints. As we shall see later, this last $r$-matrix is
still present
in the Poisson brackets of the currents. Finally, notice that in
eq.(\ref{eq31})
the group elements are on the left of $r_m$ and, as usual, on the right of
$r_c$. As a consequence, this Poisson bracket  is not invariant under arbitrary
right or left translations.

The verification of the  Jacobi identity  turns out  not to require the use of
the constraints (\ref{eq17}) on $G$, or the use of $r_{m}^{2} =1$. Indeed the
Jacobi identity just requires $r_{m}$ and $r_{c}$ to be solutions of the
modified Yang-Baxter equation.
More precisely in the sum of circular permutations the terms containing both
$r_{m}$ and $r_{c}$ vanish. Then, the terms containing $r_{m}$ only give, using
the modified Yang-Baxter equation
\beq
\frac{1}{4} G_{1X} G_{2Y} G_{3Z} [C_{12} , C_{13} ] \delta^{2}_{XY}
\delta^{2}_{XZ}. \label{eq47}
\eeq
In contrast to the bosonic and $N=1$ cases, the sum of the terms containing
$r_{c}$ and/or $C$ does not vanish. It gives as in eq.(\ref{eq95})
\beq
\frac{1}{4} (\varepsilon^{2}_{XY} \varepsilon^{2}_{XZ} +
\varepsilon^{2}_{YZ} \varepsilon^{2}_{YX}+
\varepsilon^{2}_{ZX} \varepsilon^{2}_{ZY} -1 ) [ C_{12} , C_{13} ] G_{1X}
G_{2Y} G_{3Z},
\eeq
which, using the property (\ref{eq30}) of the $(2,0)$ sign distribution,
becomes
\beq
- \frac{1}{4} [ C_{12}, C_{13} ] G_{1X} G_{2Y} G_{3Z} \delta^{2}_{XY}
\delta^{2}_{XZ}. \label{eq48}
\eeq
Since $C$ is the Casimir operator, the sum of the two terms in eq.(\ref{eq47})
and eq.(\ref{eq48}) vanishes.

	\subsection{Poisson brackets of the currents}

We can now compute the Poisson bracket of the currents. If ${\cal A}(X)$ is
some Lie algebra valued superfield, we define its ``covariant derivative'' by
\beq \dd{\cal A}=G^{-1}D(G{\cal A}G^{-1})G
=D{\cal A}+[J, {\cal A}]
\eeq
Since $\rmt_{12} = 2 \rmt_{12}^{\pm} \mp C_{12}$ we may rewrite (\ref{eq31}) as
\bean
\{ G_{1X}, G_{2Y} \}  =  G_{1X} G_{2Y} \rmt_{12}^{\pm} \delta^{2}_{XY}
+ \frac{1}{2} \left[\rc_{12} + C_{12} (\varepsilon^{2}_{XY} \mp
\delta^{2}_{XY})\right]G_{1X} G_{2Y}.
\eean
Using the property eq.(\ref{eqq7}) of the sign distribution, one finds
\bea
\{ J_{1X} , G_{2Y} \} &=& G_{2Y} \dd_{1X}({\rmt}_{12}^{+} \dexyd), \\
\{ J_{1X} , J_{2Y} \}  &=& - \dd_{2Y}\dd_{1X}({\rmt}_{12}^{+} \dexyd),
\label{eq80} \\
\{ J_{1X} , { \bar J}_{2Y} \}  &=& - \ddd_{2Y}\dd_{1X}({\rmt}_{12}^{+} \dexyd),
\label{eq81} \\
\{ { \bar J}_{1X} , { \bar J}_{2Y} \}  &=& - \ddd_{2Y}\ddd_{1X}({\rmt}_{12}^{-}
\dexyd). \label{eq82}
\eea
More explicitly, expanding the covariant derivatives in these expressions we
get
\bea
\{ J_{1X}, J_{2Y} \} &=& [ J_{1X}, \rmt_{12}^{+} ] D_{Y} \dexyd - [ J_{2Y} ,
\rmt_{12}^{+}] D_{X} \dexyd  \nonumber  \\
&&+ [J_{1Y},[J_{2Y},\rmt_{12}^{+}]] \dexyd,\label{eq49} \\
\{ J_{1X} , {\bar J}_{2Y} \} &=& [J_{1X} , \rmt_{12}^{+} ] {\bar D}_{Y} \dexyd
- [{\bar J}_{2Y}, \rmt_{12}^{+} ] D_{X} \dexyd \nonumber \\
&&+ [ J_{1Y}, [ {\bar J}_{2Y}, \rmt_{12}^{+} ]] \dexyd - \rmt_{12}^{+} {\bar
D}_{Y} D_{X} \dexyd. \label{eq83}
\eea
Notice that these Poisson brackets are by construction compatible with the
non-linear constraints (\ref{eqq2}). For what concerns the linear
constraints (\ref{eq44}), although they are not needed to verify the Jacobi
identity, they are also compatible with the Poisson brackets. This means that
the Poisson bracket of any function of the phase space variables $G$ with the
constraints vanishes when the constraints are applied.
Note finally that there is no central term in the Poisson bracket of two $J's$
since $D^{2} =0$.

In order to recover the results of \cite{hull90-2}, we need to suppose that the
currents are constrained by (\ref{eq44}), and also that $r_{m}^{2} =1$. We
denote by $T_{a}$, (respectively by $T_{\bar a}$), a basis of ${\cal G}_{+}$,
(${\cal G}_{-}$). Then  we have $\rmt_{12}^{+} = T^{\bar e}\otimes T_{\bar e} =
T_{e} \otimes T^{e}$. The linear constraints (\ref{eq44}) then translate into
$J=J^aT_a$, $\bar J=\bar J^{\bar a}T_{\bar a}$. The non-linear constraints
(\ref{eqq2}) become
\beq
DJ^{a} = - \frac{1}{2} f^{a}_{\hspace{4pt} bc} J^{b} J^{c} \esp {\bar D}{\bar
J}^{\bar a} = - \frac{1}{2} f^{\bar a}_{\hspace{4pt} {\bar b}{\bar c}} {\bar
J}^{\bar b} {\bar J}^{\bar c}
\eeq
which correspond to the constraints in \cite{hull90-2}.
Then, from eqs.(\ref{eq49},\ref{eq83}), we find
\bea
\{ J^{a}_{X} , J^{b}_{Y} \} &=& - f^{ab}_{\hspace{8pt}c} J_{Y}^{c} D_{X} \dexyd
+ f^{a}_{\hspace{4pt}ec} f^{be}_{\hspace{8pt}d} J_{Y}^{c} J_{Y}^{d} \dexyd,
\label{eq84} \\
\{ J^{a}_{X}, {\bar J}^{\bar b}_{Y} \} &=& - g^{a {\bar b}}
{\bar D}_{Y} D_{X} \dexyd + f^{a{\bar b}}_{\hspace{8pt}c} J_{Y}^{c} {\bar
D}_{X} \dexyd \nonumber \\
& -& f^{ab}_{\hspace{8pt} \bar c} {\bar J}_{Y}^{\bar c} D_{X} \dexyd + f^{\bar
b}_{\hspace{4pt}\bar ce} f^{a {\bar c}}_{\hspace{8pt}d} J_{Y}^{d} {\bar
J}_{Y}^{\bar e} \dexyd, \label{eq85}
\eea
which correspond to the classical limit of the operator product expansions
derived in \cite{hull90-2}.

There exists another interesting way of writing the Poisson brackets of the
currents. Again we do not suppose neither that the currents are constrained nor
that $r_{m}^{2}=1$.
For any functional $F$ of the currents we define the functional derivative with
respect to $J$ as
\beq
\delta F \equiv \int dY_{c} \tr \frac{\delta F}{\delta J_{Y}} \delta J_{Y}
\,\,\,\,\,\, {\mathrm{and}}\,\,\,\,\,\,
{\cal D}_{Y} (\frac{\delta F}{\delta J_{Y}}) = 0
\label{eqq4}\eeq
The integration is over the chiral superspace $V^{c}$. With such a definition,
the Grassmann parity of the derivative ${\delta F}\over{\delta J_{X}}$ is the
same as that of the functional $F$.
We have in particular
\beq
\frac{\delta J^{a}(X)}{\delta J_{Y}} = {\cal D}_{Y} \left( T^{a} \dexyd
\right).
\eeq
Similar definitions may be given for the functional derivative with respect to
$\bar J$.
The Poisson brackets (\ref{eq80}-\ref{eq82}) may then be written for two
functionals $F(J,\bar J)$ and $H(J,\bar J)$ as
\beq
\{ F(J,\bar J), H(J,\bar J) \} = \int dY \tr \left[ \frac{\delta F}{\delta
J_{Y}} r_{m}^{+}(\frac{\delta H}{\delta J_{Y}}+
\frac{\delta H}{\delta {\bar J}_{Y}}) +
\frac{\delta F}{\delta {\bar J}_{Y}} r_{m}^{-}(
\frac{\delta H}{\delta J_{Y}}+
\frac{\delta H}{\delta {\bar J}_{Y}})\right].
\eeq
It is interesting to study the properties of the Poisson bracket in this
formalism. In order to demonstrate the graded antisymmetry, we need the
identity
\beq
\tr\, D_Y\left(\frac{\delta F}{\delta J_{Y}}\frac{\delta H}{\delta
J_{Y}}\right)=0
\eeq
which follows easily from the definition eq.(\ref{eqq4}) of the functional
derivative. Then we use the antisymmetry of the r-matrix $r_m$, which leads to
$\tr(A r_m^+(B))=-\tr(r_m^-(A)B)$. The Jacobi identity follows from the
modified Yang-Baxter equation of $r_{m}$ and from the property:
\beq
\left[ \frac{\delta}{\delta J_{1Y}}, \frac{\delta}{\delta J_{2Z}} \right] = -
\left[ C_{12}, \frac{\delta}{\delta J_{1Y}} \right] \delta^{2}_{YZ}
\eeq
that can be proved for example on $J^{a}(X)$.

	\subsection{$N=2$ superconformal generator}

The infinitesimal action of the $N=2$ superconformal group on the phase space
variables $G$ is
\beq
\delta G=(\kappa\partial+D\kappa\bar D+\bar D\kappa D)G,\ \ \
\partial_{--}\kappa=0.
\label{eqq5}\eeq
This action preserves the constraints eq.(\ref{eq44}). It is known to be
generated through the Poisson bracket (\ref{eq31}) by the superconformal
tensor $T = \tr (J^{(+)} {\bar J}^{(-)})$. Here we shall show a little
more, which is that $T$ also defines an action of the superconformal group in
the extended phase space obtained by relaxing the constraints eq.(\ref{eq44}).
However, this action differs from eq.(\ref{eqq5}).
We give below the general idea of the proof. Let us consider the
transformations
\beq
\delta_{\kappa} G_{X} \equiv \tr_{1} \int dY \kappa_{Y} \{ J_{1Y}^{(+)} {\bar
J}_{1Y}^{(-)} , G_{2X} \}
\eeq
As a first step we find
\beq
G^{-1} \delta_{\kappa} G = {\bar D} (\kappa J^{(+)}) - D(\kappa {\bar J}^{(-)})
+ \kappa \{J^{(+)}, {\bar J}\}^{(+)} + \kappa \{ J, {\bar J}^{(-)} \}^{(-)},
\eeq
which we can write as
\beq
G^{-1} \delta_{\kappa} G = ({\bar D} \kappa J + D \kappa {\bar J} + \kappa
G^{-1} \partial G ) + \beta_{\kappa}. \label{eq107}
\eeq
The first term corresponds to the $N=2$ superconformal  transformation
(\ref{eqq5}) of a scalar superfield. The supplementary term
\beq
\beta_{\kappa} \equiv {\bar D} (\kappa J^{(-)}) - D (\kappa {\bar J}^{(+)}) +
\kappa \{ J^{(-)} , {\bar J}^{(+)} \} \label{eq88}
\eeq
vanishes in the constrained theory.
After some algebra we find for the variations of $J^{(+)}$ and ${\bar J}^{(-)}$
\bea
\delta_{\kappa} J^{(+)} &=& (D\kappa {\bar D} J^{(+)} + {\bar D} \kappa D
J^{(+)} + D {\bar D} \kappa J^{(+)} + \kappa \partial J^{(+)} ) + \mu_{\kappa},
\\
\delta_{\kappa} {\bar J}^{(-)} &=& (D \kappa {\bar D} {\bar J}^{(-)} + {\bar D
} \kappa D {\bar J}^{(-)} + {\bar D} D \kappa {\bar J}^{(-)} + \kappa \partial
{\bar J}^{(-)}) + {\bar \mu}_{\kappa},
\eea
where
\bea
\mu_{\kappa} &\equiv& [ J^{(+)}, \beta_{\kappa} ]^{(+)} + \kappa [ \{ J^{(+)},
J^{(-)} \}^{(-)} , {\bar J}^{(+)} ]^{(+)}, \\
{\bar \mu}_{\kappa} &\equiv& - [{\bar J}^{(-)}, \beta_{\kappa}]^{(-)} - \kappa
[ J^{(-)}, \{ {\bar J}^{(+)}, {\bar J}^{(-)}\}^{(+)} ]^{(-)}.
\eea
We can easily check that in the variation of $T$ the sum
$\mathrm{tr}\mu_{\kappa} {\bar J}^{(-)} +\mathrm{tr} J^{(+)} {\bar
\mu}_{\kappa} $ vanishes and therefore we find the usual transformation of the
superconformal tensor
\beq
\delta_{\kappa} T = \partial \kappa T + D \kappa {\bar D} T + {\bar D} \kappa D
T + \kappa \partial T.
\eeq
Taking as the Hamiltonian
$H \equiv \int dX\,\, T$,
the equation of motion in the constrained theory is $\partial_{--} G = 0$.

	\subsection{Loop group and Poisson-Lie symmetries}

We now show that the loop group transformations described in (\ref{eq62})
are generated by ${\cal J} \equiv D [ (\theta - {\bar \theta}) J ]$ and $ {\bar
{\cal J}} \equiv {\bar D} [ (\theta - {\bar \theta}) {\bar J} ]$.
Let us concentrate on ${\cal J}$, which is a chiral superfield and verifies
${\cal J}^{(-)} = 0$. We consider the charge
\beq
Q[\omega^{(-)}]= \int dX_{c} \,\mathrm{tr}\, {\omega^{(-)}}_{X}  {\cal J}_{X},
\eeq
where $\omega^{(-)}$ is a chiral superfield, $ D\omega^{(-)}=0$,
satisfying moreover $\partial_{--}\omega^{(-)}=0$.
It generates the variation $\delta G_X=\{ Q[\omega^{(-)}],G_X\}$
which may be written successively
\bea
\delta G_{X} &=&  \tr_{1} \int dY_{c}   D_{Y} [ {\omega^{(-)}}_{1Y_{c}} (\eta -
{\bar \eta}) \{ J_{1Y}, G_{2X} \}], \\
&=& \tr_{1} \int dY {\omega^{(-)}}_{1Y_{c}} (\eta - {\bar \eta}) \{ J_{1Y},
G_{2X} \}, \\
&=&  G_{X} \tr_{1} \int dY {\omega^{(-)}}_{1Y_{c}} (\eta - {\bar \eta})
\dd_{1Y}({\rmt}_{12}^{+} \dexyd).
\eea
After integration by parts, we finally obtain
\beq
\delta G = G(  \omega^{(-)} - (\theta - {\bar \theta}) r_{m}^{-} [ J ,
\omega^{(-)} ] )
\eeq
which is identical to eq.(\ref{eq62}).
Let us check that this transformation is a symmetry of the $N=2$ chiral WZNW
model. We compute the variation of the hamiltonian $H$.
 Under a general transformation $\delta G = G \Omega$ we have
\beq
\delta H = \int dX \tr (D\Omega^{(+)} {\bar J}^{(-)} + J^{(+)} {\bar D}
\Omega^{(-)}).
\eeq
In the particular case $\Omega = \omega^{(-)} - (\theta - {\bar \theta})
r_{m}^{-} [ J , \omega^{(-)} ] $, it becomes after a short calculation
\beq
\delta H =  \int dX \tr {\bar D}\omega^{(-)} {\cal J} .\\
\eeq
The last step is to extract one spinor derivative $D$ from the measure,
and use $D\bar D\omega^{(-)}=\partial\omega^{(-)}={\partial\over\partial
t}\omega^{(-)}$
to write
\beq
\delta H =  \int dX_c \tr ({\partial \over\partial t}\omega^{(-)} ){\cal J} ,\\
\eeq
which is equivalent to the conservation of the charge $Q[\omega^{(-)}]$.

 Since $G$ and its monodromy have the same Poisson brackets as in the bosonic
and $N=1$ cases, the $N=2$ chiral WZNW model admits a Poisson-Lie symmetry
corresponding to the left action of $G_{PL}$. It is still generated by the
monodromy.
If we now consider the model {\em without} the constraints eq.(\ref{eq44})
on the currents, there exists also a right action $G(X) \rightarrow G(X)h(X)$
of the infinite Poisson-Lie group with the Poisson structure
\beq
\{h_{1X}, h_{2Y} \} \equiv \left[h_{1X}h_{2Y}, {r_{m}}_{12} \right] \dexyd .
\eeq
This is just the Sklyanin bracket induced by $r_{m}$, extended to $N=2$
superfields.

\section*{Conclusion}

This article was originally conceived as a small step towards a better
understanding of the geometry and symmetries of supersymmetric
non-linear sigma models
with torsion, using the important example of the WZNW model.
Although some of the original goals such as an
explicitly supersymmetric action principle have not been fulfilled,
we met with the interesting fact that a structure typical of integrable
systems, namely an r-matrix, appears in this model as a consequence
of a supersymmetry requirement. It seems reasonable to hope that some of the
techniques developped in the context of integrable models should prove useful
here as well. We intend to continue this study, in particular in the quantum
case.
\vspace*{1.0cm}

{{\bf Acknowledgements}\\
We thank Laurent Freidel for many useful discussions.}
\vspace*{1.0cm}
\section*{Appendix}
\label{eq104}
\renewcommand{\theequation}{\Alph{section}.\arabic{equation}}
\setcounter{equation}{0}
\setcounter{section}{1}
In this appendix we define the generalizations of the Dirac and sign
distributions for the $(1,0)$ and $(2,0)$ superspaces. We denote respectively
by $\delta^{i}$ and $\varepsilon^{i}$ the Dirac and sign distributions for the
$(i,0)$ superspace. Let us recall first that the bosonic sign distribution
satisfies the quadratic  equation
\beq
\varepsilon^{0}_{xy} \varepsilon^{0}_{xz} + \varepsilon^{0}_{yz}
\varepsilon^{0}_{yx}  + \varepsilon^{0}_{zx} \varepsilon^{0}_{zy}  = 1.
\label{eq25}
\eeq
The $(1,0)$-Dirac distribution is $\dexyu \equiv \dexyz (\theta^{+}_{1} -
\eta^{+}_{1})$. The $(1,0)$ sign distribution $\varepsilon^{1}$ is required to
be invariant under supersymmetry, and to satisfy the equations
\beq
{D_{+}^{1}}_{\mathbf{x}} \epu = 2\dexyu ,\,\,\,\,\,
\epu = - \enu \label{eq33}
\eeq
It is easily shown that this system determines $\varepsilon^{1}$ to be
\beq
\epu = {D_{+}^{1}}_{\mathbf{x}} (\epz (\theta^{+}_{1} - \eta^{+}_{1})).
\label{eq39}
\eeq
It satisfies the following properties
\bea
{\epu}|_{\theta^{+}_{1},\eta^{+}_{1} =0} &=& \epz,\\
\varepsilon^{1}_{\mathbf{xy}} \varepsilon^{1}_{\mathbf{xz}}  +
\varepsilon^{1}_{\mathbf{yz}} \varepsilon^{1}_{\mathbf{yx}}  +
\varepsilon^{1}_{\mathbf{zx}} \varepsilon^{1}_{\mathbf{zy}}  &=& 1.
\label{eq27}
\eea
The $(2,0)$-Dirac distribution is defined as
\beq
\dexyd \equiv  \dexyz (\theta^{+}-\eta^{+}) ({\bar \theta^{+}}-{\bar \eta^{+}})
= -2 \dexyz (\theta_{1}^{+}-\eta_{1}^{+}) (\theta_{2}^{+}-\eta_{2}^{+})
\label{eq58}
\eeq
such that
$\int dx d{\bar \theta}^{+} d\theta^{+} \dexyd  f(x,\theta^{+},{\bar
\theta}^{+}) = f(y,\eta^{+},{\bar \eta}^{+})$ and $\dexyd = \delta^{2}_{YX}$.
The $(2,0)$ sign distribution is required to be invariant under supersymmetry,
and to solve the system of equations
\beq
{D_{+}^{1}}_{X} \epd ={D_{+}^{2}}_{X} \dexyd,\,\,\,{D_{+}^{2}}_{X} \epd
={D_{+}^{1}}_{X} \dexyd, \,\,\,
\epd = - \enpd . \label{eq28}
\eeq
As in the $(1,0)$ case this system fixes $\varepsilon^{2}$ to be
\beq
\epd = \frac{1}{2} [{{\overline D}_+}_{X},{D_{+}}_{X} ]( \epz
(\theta^{+}-\eta^{+}) ({\bar \theta^{+}}-{\bar \eta^{+}}))
\eeq
It verifies the properties:
\bea
\epd|_{\theta^{+}_{1},\eta^{+}_{1} =0} &=& \epu,\\
\varepsilon^{2}_{XY} \varepsilon^{2}_{XZ} +  \varepsilon^{2}_{YZ}
\varepsilon^{2}_{YX}  + \varepsilon^{2}_{ZX} \varepsilon^{2}_{ZY} &=& 1 -
\delta^{2}_{XY} \delta^{2}_{XZ}. \label{eq30}\\
{{\overline D}_+}_{X}(\epd+\dexyd)=0, &\,\,\,\, &{D_{+}}_{X}(\epd-\dexyd)=0.
\label{eqq7}\eea

\end{document}